\documentclass{desyproc}

\begin{document}
\title{The QCD critical end point driven by an external magnetic field 
in asymmetric quark matter}

\author{{\slshape Pedro Costa$^1$, M\'arcio Ferreira$^1$, Constan\c ca Provid\^encia$^1$, 
Hubert Hansen$^2$, D\'ebora P. Menezes$^2$}\\[1ex]
$^1$Centro de F\'isica Computacional, Department of Physics, University of Coimbra,
P-3004-516 Coimbra, Portugal\\
$^2$IPNL, Universit\'e de Lyon/Universit\'e Lyon 1, CNRS/IN2P3, 4 rue E.Fermi,
F-69622 Villeurbanne Cedex, France\\
$^2$Departamento de F\'isica, CFM, Universidade Federal de Santa Catarina,
Florian\'opolis, SC, CP 476, CEP 88.040-900, Brazil }

\contribID{75}

\confID{8648}  
\desyproc{DESY-PROC-2014-04}
\acronym{PANIC14} 
\doi  

\maketitle

\begin{abstract}
The effect of the isospin/charge asymmetry and an external magnetic field in 
the location of the critical end point (CEP) in the QCD phase diagram is 
investigated.
By using the 2+1 flavor Nambu--Jona-Lasinio model with Polyakov loop (PNJL), 
it is shown that the isospin asymmetry shifts the CEP to larger baryonic 
chemical potentials and smaller temperatures, and in the presence of a large 
enough isospin asymmetry the CEP disappears.
Nevertheless, a sufficiently high external magnetic field can drive the system 
into a first order phase transition again.
\end{abstract}

The QCD phase diagram under extreme conditions of density, temperature and 
magnetic field is the subject of intense studies \cite{Marcio}.
Understanding the effect of an external magnetic field on the structure of the 
QCD phase diagram is very important: these extremely strong magnetic fields are 
expected to affect the measurements in heavy ion collisions (HIC) at very high 
energies, to influence the behavior of the first stages of the Universe and are 
also relevant to the physics of compact astrophysical objects like magnetars.

On the other hand, the effect of the isospin/charge asymmetry in the QCD phase 
diagram is also very interesting due to its role on the location of the the 
critical end point (CEP): it was shown that for a sufficiently asymmetric 
system the CEP is not present \cite{ueda13,abuki13}.

In the present work we describe quark matter subject to strong magnetic fields 
within the 2+1 PNJL model. 
The PNJL Lagrangian with explicit chiral symmetry breaking where 
the quarks couple to a (spatially constant) temporal background gauge field, 
represented in terms of the Polyakov loop and in the presence of an external 
magnetic field is given by \cite{PNJL}:
\begin{eqnarray}
{\cal L} = {\bar{q}} \left[i\gamma_\mu D^{\mu}-{\hat m}_f \right ] q ~+~ 
	G \sum_{a=0}^8 \left [({\bar q} \lambda_ a q)^2 + ({\bar q} i\gamma_5 \lambda_a q)^2 \right ]\nonumber\\
	-K\left\{{\rm det} \left [{\bar q}(1+\gamma_5) q \right] + 
	{\rm det}\left [{\bar q}(1-\gamma_5)q\right]\right\} + 
	\mathcal{U}\left(\Phi,\bar\Phi;T\right) - \frac{1}{4}F_{\mu \nu}F^{\mu \nu},
	\label{Pnjl}
\end{eqnarray}
where the quark sector is described by the  SU(3) version of the
Nambu--Jona-Lasinio model which includes the scalar-pseudoscalar 
(chiral invariant) and the t'Hooft six fermion interactions that breaks the 
axial $U_A(1)$ symmetry.
The $q = (u,d,s)^T$ represents a quark field with three flavors, 
${\hat m}_f= {\rm diag}_f (m_u^0,m_d^0,m_s^0)$ is the corresponding (current) 
mass matrix, { $\lambda_0=\sqrt{2/3}I$  where $I$ is the unit matrix in the 
three flavor space, and $0<\lambda_a\le 8$ denote the Gell-Mann matrices.
The coupling between the magnetic field $B$ and quarks, and between the 
effective gluon field and quarks is implemented  {\it via} the covariant 
derivative 
$D^{\mu}=\partial^\mu - i q_f A_{EM}^{\mu}-i A^\mu$ 
where $q_f$ represents the quark electric charge ($q_d = q_s = -q_u/2
= -e/3$),  $A^{EM}_\mu$ and 
$F_{\mu \nu }=\partial_{\mu }A^{EM}_{\nu }-\partial _{\nu }A^{EM}_{\mu }$ 
are used to account for the external magnetic field and 
$A^\mu(x) = g_{strong} {\cal A}^\mu_a(x)\frac{\lambda_a}{2}$ where
${\cal A}^\mu_a$ is the SU$_c(3)$ gauge field.
We consider a  static and constant magnetic field in the $z$ direction, 
$A^{EM}_\mu=\delta_{\mu 2} x_1 B$.
In the Polyakov gauge and at finite temperature the spatial components of the 
gluon field are neglected: 
$A^\mu = \delta^{\mu}_{0}A^0 = - i \delta^{\mu}_{4}A^4$. 
The trace of the Polyakov line defined by
$ \Phi = \frac 1 {N_c} {\langle\langle \mathcal{P}\exp i\int_{0}^{\beta}d\tau\,
A_4\left(\vec{x},\tau\right)\ \rangle\rangle}_\beta$
is the Polyakov loop which is the {\it exact} order parameter of the 
$\mathbb{Z}_3$ symmetric/broken phase transition in pure gauge.

To describe the pure gauge sector an effective potential 
$\mathcal{U}\left(\Phi,\bar\Phi;T\right)$ is chosen in order to reproduce 
the results obtained in lattice calculations \cite{Ratti:2006}:
\begin{eqnarray}
	& &\frac{\mathcal{U}\left(\Phi,\bar\Phi;T\right)}{T^4}
	= -\frac{a\left(T\right)}{2}\bar\Phi \Phi
	+ b(T)\mbox{ln}\left[1-6\bar\Phi \Phi+4(\bar\Phi^3+ \Phi^3)-3(\bar\Phi \Phi)^2\right],
	\label{Ueff}
\end{eqnarray}
where $a\left(T\right)=a_0+a_1\left(\frac{T_0}{T}\right)+a_2\left(\frac{T_0}{T}\right)^2$, 
$b(T)=b_3\left(\frac{T_0}{T}\right)^3$.
The standard choice of the parameters for the effective potential $\mathcal{U}$ 
is $a_0 = 3.51$, $a_1 = -2.47$, $a_2 = 15.2$, and $b_3 = -1.75$.
$T_0$ is the critical temperature for the deconfinement phase transition within 
a pure gauge approach: it was fixed to a constant $T_0=270$ MeV, according to 
lattice findings.
The parameters of the model are $\Lambda = 602.3 \, {\rm MeV}$, 
$m_u^0= m_d^0=\,  5.5 \,{\rm MeV}$, $m_s^0=\,  140.7\, {\rm MeV}$, 
$G \Lambda^2= 1.385$ and $K \Lambda^5=12.36$. 

The thermodynamical potential for the three flavor quark sector, $\Omega$, 
in the mean field approximation is written as
\begin{eqnarray}
	\Omega(T,\,B,\,\mu_f)= 2G \sum_{f=u,\,d,\,s} \left\langle \bar{q}_fq_f\right\rangle^2 
	-4K \, \left\langle \bar{q}_uq_u\right\rangle \left\langle \bar{q}_dq_d\right\rangle
	\left\langle \bar{q}_sq_s\right\rangle
	+\left(\Omega_f^{vac}+\Omega_f^{mag}+\Omega_f^{med}\right),
	\label{Omega}
\end{eqnarray}
where the vacuum $\Omega_f^{vac}$, the magnetic $\Omega_f^{mag}$, 
the medium contributions $\Omega_f^{med}$ and the quark condensates 
$\left\langle \bar{q}_fq_f\right\rangle$ have been evaluated with 
great detail in \cite{prc,hotnjl}. 
The mean field equations are obtained by minimizing the thermodynamical 
potential (\ref{Omega}) with respect to the order parameters
$\left\langle \bar{q}_fq_f\right\rangle$, $\Phi$ and $\bar{\Phi}$.

\begin{figure}[t]
\centerline{\includegraphics[width=0.5\textwidth]{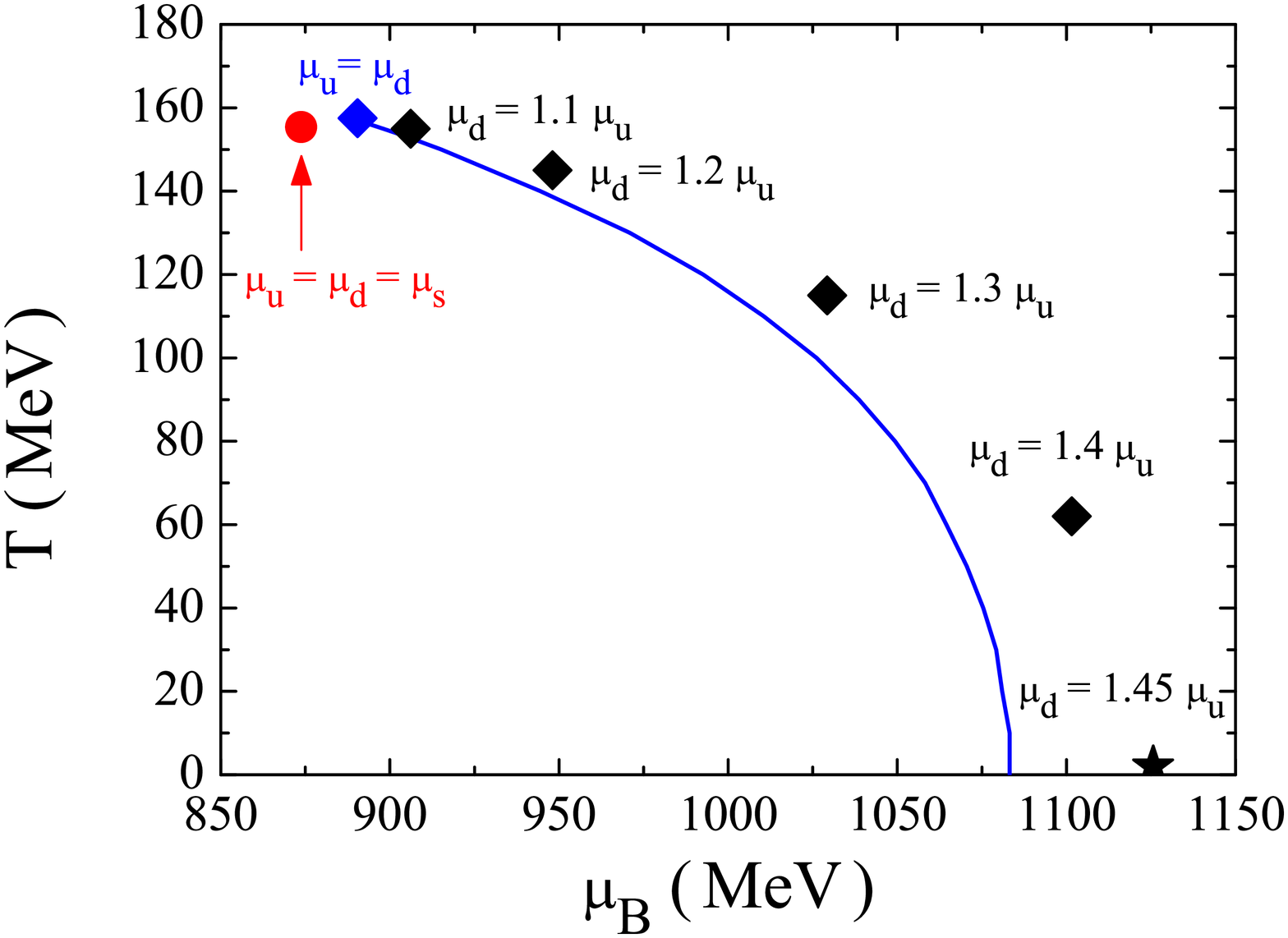}
\includegraphics[width=0.5\textwidth]{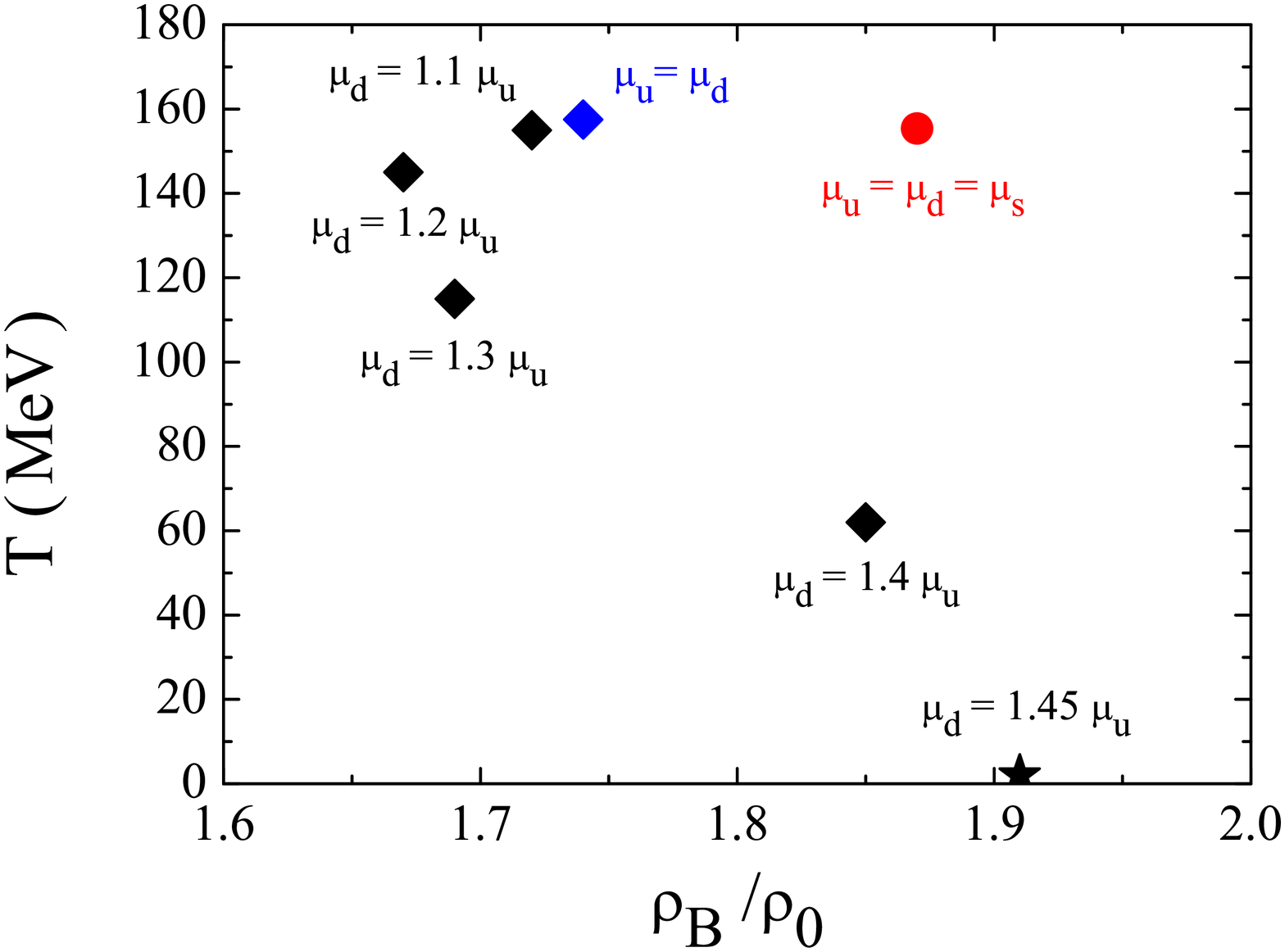}}
\caption{
The influence of the isospin in the location of the CEP within the PNJL model: 
the full line is the first order phase transition line for zero isospin matter 
($\mu_u=\mu_d,\, \mu_s = 0$). The chemical potential for the strange quark is 
always taken equal to zero, except the for the red point ($\mu_u=\mu_d=\mu_s$) 
which is given for reference. 
When $\mu_d > 1.45\mu_u$ the CEP doesn't exist anymore. 
}
\label{Fig:1}
\end{figure}

We start the discussion of our results by the location of the CEP when no 
external magnetic field is present.

It has been shown that the location of the CEP depends on the isospin 
\cite{Costa:2013zca}: as an example, in $\beta-$equilibrium matter the CEP 
occurs at larger baryonic chemical potentials and smaller temperatures \cite{Costa:2013zca}. 
Indeed, we are interested in $d$-quark rich matter as it occurs in neutron 
stars and in HIC: isospin asymmetry in neutron matter has 
$\mu_d \sim 1.2 \mu_u$, and presently the attained isospin asymmetry in HIC 
corresponds to $\mu_u<\mu_d< 1.1 \mu_u$. 
In the present work the effect of isospin on the CEP is studied:
we increase systematically $\mu_d$ with respect to $\mu_u$ taking the $s$-quark 
chemical potential equal to zero ($\mu_s=0$ leads to all CEP's occur at 
$\rho_s=0$). 

The results for the CEP in the previous conditions are presented in 
Fig. \ref{Fig:1}. For reference we also show the red full point that 
corresponds to the CEP with $\mu_u=\mu_d=\mu_s$. 
When the isospin asymmetry is increased the CEP moves to smaller temperatures 
and larger baryonic chemical potentials. 
\begin{figure}[t]
\centerline{\includegraphics[width=0.5\textwidth]{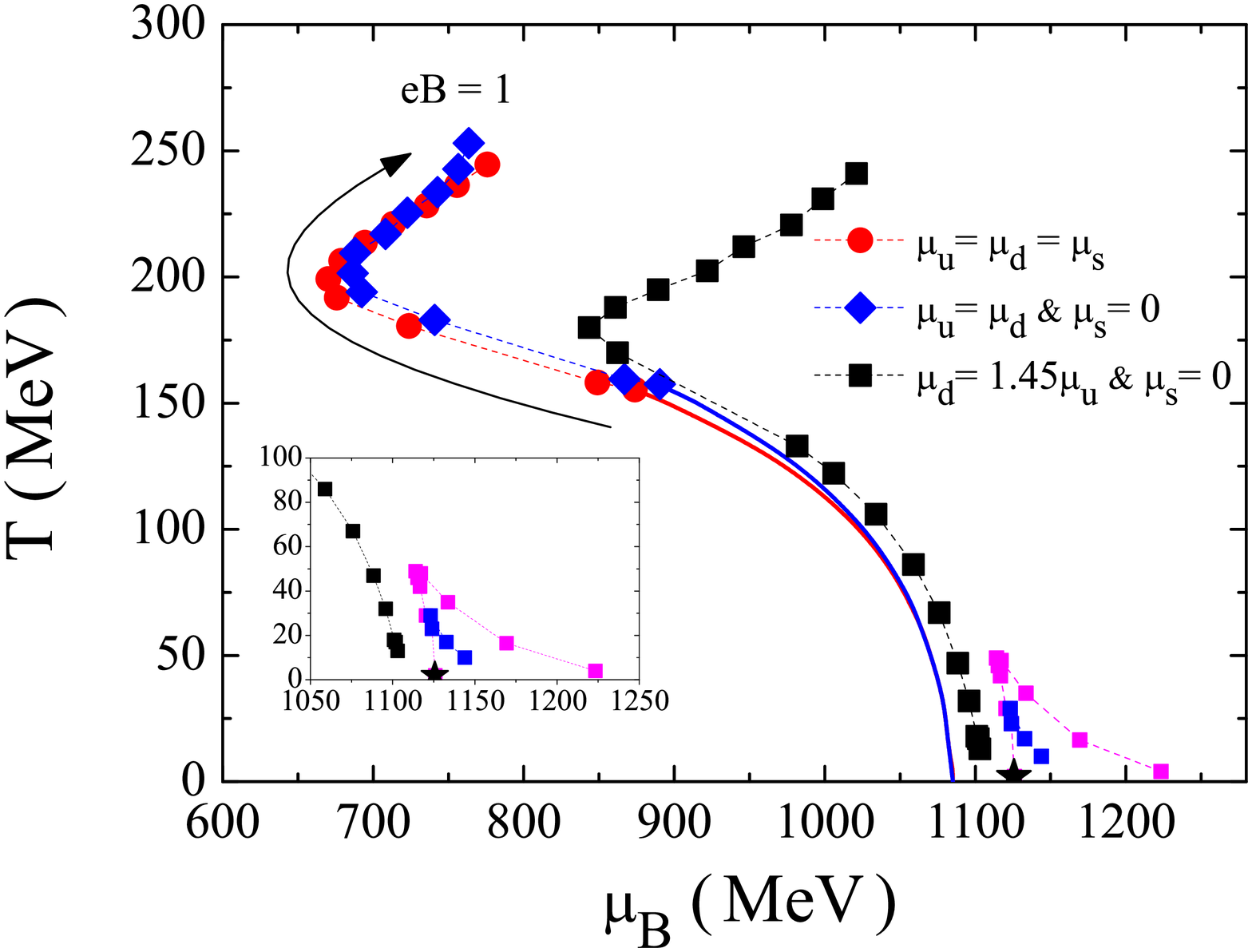}
\includegraphics[width=0.5\textwidth]{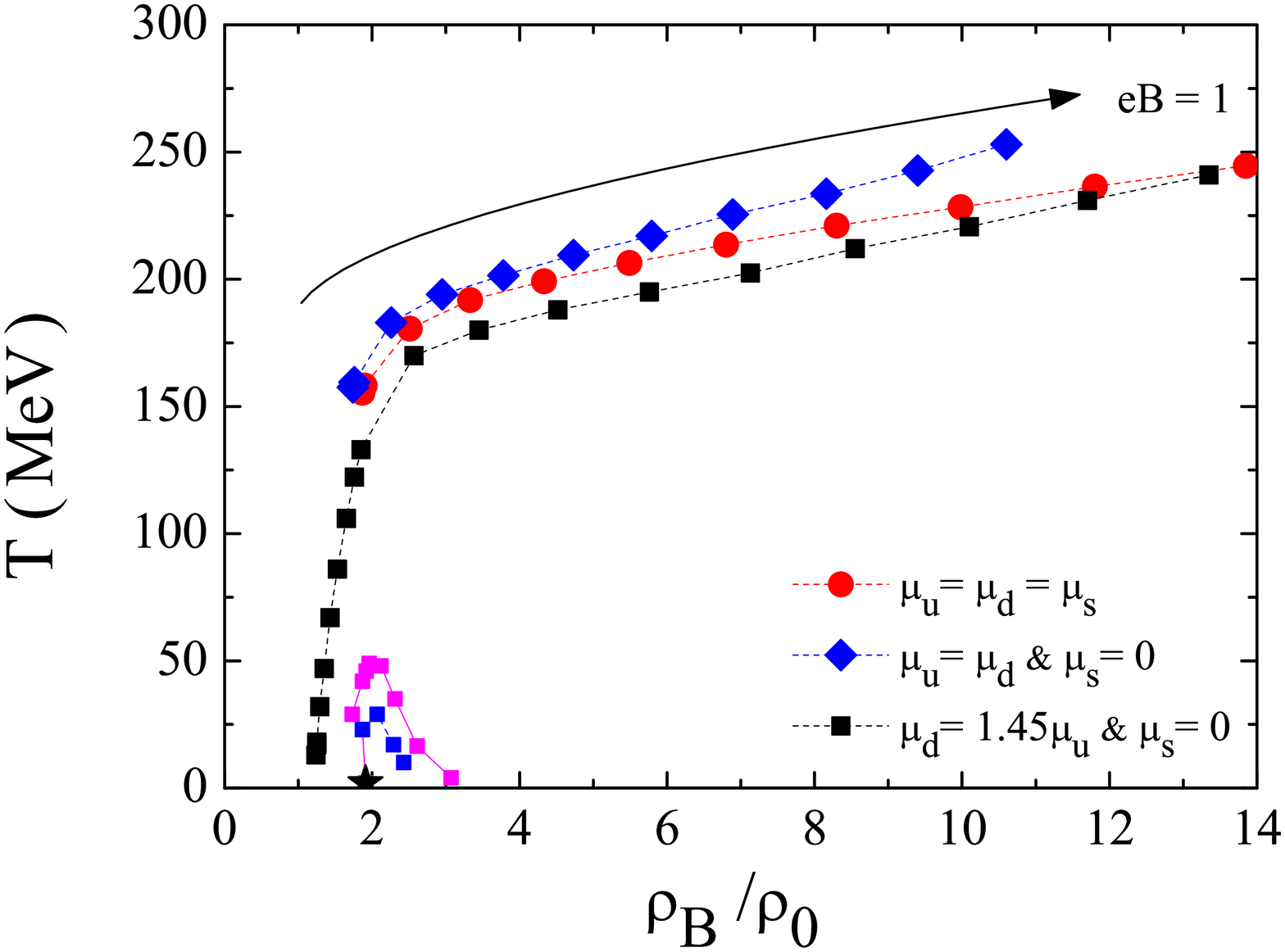}}
\caption{
Effect of an external magnetic field on the location of the CEP: 
$T^{CEP}$ vs $\mu_B^{CEP}$ (left panel) and $T^{CEP}$ vs $\rho_B^{CEP}$ (right panel).
The full lines correspont to the first order transitions at $eB=0$.
Three scenarios are shown: $\mu_u=\mu_d=\mu_s$ (red dots), $\mu_u=\mu_d;\,\mu_s=0$ 
(blue diamonds) and $\mu_d=1.45 \mu_u$, $\mu_s=0$ (black squares) 
corresponding to the threshold isospin asymmetry above which no CEP occurs. 
In the last scenario for strong enough magnetic fields and low temperatures
two or more CEP exist at different temperatures for a given magnetic field 
intensity (pink and blue squares).
}
\label{Fig:2}
\end{figure}

When the asymmetry is large enough, $\mu_d=1.45\mu_u$, the CEP disappears 
(this CEP is represented in Fig. \ref{Fig:1} by a star at $T=0$). 
This scenario leads to $|\mu_u-\mu_d|=|\mu_I|=|\mu_Q|=130$ MeV, below the pion 
mass and, accordingly, no pion condensation occurs under these conditions. 

The CEP for ($T,\rho_B$) plane is shown in the right panel of 
Fig. \ref{Fig:2}. When $\mu_u<\mu_d< 1.2 \mu_u$ the baryonic density of the CEP 
decreases with asymmetry but for $\mu_d \gtrsim 1.2 \mu_u$ the opposite occurs 
and at the threshold ($\mu_d=1.45\mu_u$) $\rho_B\sim 1.91 \rho_0$.

Now, we investigate how a static external magnetic field will influence the 
localization of the CEPs  previously calculated. The results are plotted in 
Fig. \ref{Fig:2}. In the left panel of Fig. \ref{Fig:2} the red dots correspond 
to symmetric matter ($\mu_u=\mu_d=\mu_s$) and reproduce qualitatively the 
results previously obtained within the NJL model \cite{avancini2012} being the 
trend qualitatively similar: the increasing of the intensity of the magnetic 
field leads to an increase of the CEP's temperature and to a decrease of 
the CEP's baryonic chemical potential until the critical value 
$eB\sim 0.4$ GeV$^2$; for stronger magnetic fields, both $T$ and $\mu_B$ increase. 
In the right panel of Fig. \ref{Fig:2} the CEP is given in a $T$ vs. $\rho_B$
plane. The results show that when $eB$ increases from 0 to 1 GeV$^2$ the 
baryonic density at the CEP increases from 2$\rho_0$ to $14\rho_0$. 

Taking the isospin symmetric matter scenario $\mu_u=\mu_d$ and
$\mu_s=0$, the effect of the magnetic field on the CEP is very similar to
the previous one (see blue diamonds in Fig. \ref{Fig:2}): CEP's temperature 
is only slightly larger and the CEP's baryonic density is slightly smaller.

Also interesting is the case that occurs for the very asymmetric matter scenario: 
a first order phase transition driven by the magnetic field takes place
if $\mu_d\gtrsim1.45 \mu_u$. Taking the threshold value $\mu_d=1.45\mu_u$ it
is seen that for $eB<$0.1 GeV$^2$ two CEPs may appear. 
Indeed, for sufficiently small values of $eB$ the $T^{CEP}$ is small and the 
Landau level effects are visible. 

A magnetic field affects in a different way $u$ and $d$ quarks due to their 
different electric charge. A consequence is the possible appearance of
two or more CEPs for a given magnetic field intensity. 
Two critical end points occur at different values of $T$ and $\mu_B$ 
for the same magnetic field intensity for fields $0.03 \lesssim eB
\lesssim 0.07$ GeV$^2$.  Above 0.07 GeV$^2$ only one CEP remains. 
For stronger fields we get  $T^{CEP}>100$ MeV: Landau level effects are 
completely washed out at these temperatures.

\vspace{0.5cm}
{\bf Acknowledgments}

This work was supported by C.F.C., by Project No. CERN/FP/123620/2011 
developed under the initiative QREN financed by the
UE/FEDER through the program COMPETE — ``Programa
Operacional Factores de Competitividade'' by Grant
No. SFRH/BD/51717/2011, by CNPq/Brazil and by
FAPESC/Brazil.


\begin{footnotesize}


\end{footnotesize}


\end{document}